\documentclass[twocolumn,a4paper,reprint,amssymb,aps,prl,groupedaddress,superscriptaddress]{revtex4-1}
\pdfoutput=1
\usepackage{dcolumn}
\usepackage{array}
\usepackage{xcolor}
\usepackage{graphicx,bm,color,psfrag}
\usepackage{subfigure,bm}
\usepackage{amsfonts}
\usepackage{lipsum}
\usepackage{enumitem}  
\usepackage{hyperref} 
\usepackage{array}
\usepackage{resizegather}
\newcolumntype{P}[1]{>{\centering\arraybackslash}p{#1}}
\newcolumntype{M}[1]{>{\centering\arraybackslash}m{#1}}
\newcolumntype{L}{D{.}{.}{1.0}}

\def\equationeqrefname~#1\null{Eq.(#1)\null}

\newcommand{\be}{\begin{equation}}
\newcommand{\ee}{\end{equation}}

\def\D {\mbox{D}}
\def\div {\mbox{div}\,}

\def\c {\mbox{curl}\,}

\def \ts {\textstyle}

\usepackage{soul}
\begin{document}

\title{Scalar field emulator via anisotropically deformed vacuum energy:\\ Application to dark energy}

\author{\"{O}zg\"{u}r Akarsu}
\email{akarsuo@itu.edu.tr}
\affiliation{Department of Physics, Istanbul Technical University, Maslak 34469 Istanbul, Turkey}

\author{Nihan Kat{\i}rc{\i}}
\email{nihan.katirci@itu.edu.tr}
\affiliation{Department of Physics, Istanbul Technical University, Maslak 34469 Istanbul, Turkey}

\author{Anjan A. Sen}
\email{aasen@jmi.ac.in}
\affiliation{Centre for Theoretical Physics, Jamia Millia Islamia, New Delhi-110025, India.}


\author{J. Alberto V\'azquez}
\email{javazquez@icf.unam.mx}
\affiliation{Instituto de Ciencias F\'isicas, Universidad Nacional Aut\'onoma de M\'exico, Cuernavaca, Morelos, 62210, M\'exico}

\begin{abstract}
We introduce a generalization of the usual vacuum energy, called `deformed vacuum energy', which yields anisotropic pressure whilst preserving zero inertial mass density. It couples to the shear scalar in a unique way, such that they together emulate the canonical scalar field with an arbitrary potential. This opens up a new avenue by reconsidering cosmologies based on canonical scalar fields, along with a bonus that the kinetic term of the scalar field is replaced by an observable, the shear scalar. We further elaborate the aspects of this approach in the context of dark energy.

\end{abstract}
\maketitle
\emph{Introduction} --  \label{sec:intro} The $\Lambda$CDM (Lambda cold dark matter) model is established on the spatially maximally symmetric and flat Robertson-Walker (RW) spacetime, and general relativity with a positive cosmological constant. It is in good agreement with most of the currently available data \cite{Riess:1998cb,Ade:2015xua,Alam:2016hwk,Abbott:2017wau,Aghanim:2018eyx}, but suffers from theoretical issues related to $\Lambda$ \cite{Weinberg:1988cp,Sahni:1999gb,Peebles:2002gy,Padmanabhan:2002ji}. This led to a more general `dark energy' (DE) concept, for which the quintessence---described by a canonical scalar field (SF)---has been the most natural candidate \cite{Copeland:2006wr,Tsujikawa:2013fta}. It has been customary to justify the spatially flat RW background via the standard inflationary scenarios employing canonical SF \cite{Starobinsky:1980te,Guth:1980zm,Linde:1981mu,Albrecht:1982wi}, wherein the space dynamically flattens and very efficiently isotropizes (cosmic no-hair theorem \cite{Wald:1983ky,Starobinsky:1982mr,Maleknejad:2012as,Maleknejad:2012fw}). Allowing anisotropic expansion factors---while retaining isotropic spatial curvature---leads to a generalized Friedmann equation containing average Hubble parameter along with the shear scalar \cite{Collins:1972tf,Ellis:1998ct,EllisRC}. The shear scalar resembles the stiff fluid (described by an equation of state parameter $w=1$ \cite{zel61,Barrow78}, similar to the kinetic term of a canonical SF) and dilutes faster than any other physical source (for which $w=1$ is the causality limit \cite{EllisRC}) as the Universe expands. Hence, it is unlikely to have a detectable amount of anisotropic expansion in the observable Universe. Nevertheless, the interest in anisotropic cosmologies has never disappeared, as, for instance, a deviation from the stiff-fluid character of the shear scalar might imply the necessity for replacing $\Lambda$ (or SFs) by an anisotropic stress, see \cite{Barrow:1997sy} for a list. This interest has frequently been enhanced by new observations, see, for instance, \cite{Bennett11,Ade:2013kta,Schwarz:2015cma,Akrami:2019bkn,Wilczynska:2020rxx,Migkas:2020fza} and references therein, for such hints of anomalies/unexpected features in the cosmic microwave background (CMB) data from the WMAP and Planck experiments, as well as in other types of cosmological data from independent observations. And, \cite{BeltranJimenez:2007rsj,Koivisto:2005mm,Campanelli:2006vb,Koivisto:2007bp,Rodrigues:2007ny,Koivisto:2008xf,Campanelli:2007qn,Campanelli2,Campanelli:2009tk,Cea:2019gnu}
suggesting the deficiency in quadrupole moment of the CMB temperature angular power spectrum (compared to the expected from the best-fit $\Lambda$CDM) \cite{Bennett11,Ade:2013kta,Schwarz:2015cma}
can be explained via anisotropic expansion carried out well after the matter-radiation decoupling by anisotropic DE (see also \cite{Chimento:2005ua,Battye:2006mb,Koivisto:2008ig,Cooray:2008qn,Akarsu:2013dva,Koivisto:2014gia,Heisenberg:2016wtr,Yang:2018ubt,Tedesco},
and, for constraints on such models, \cite{Mota:2007sz,appleby10,Appleby:2012as,Amendola:2013qna}). Search for anisotropic expansion occupies an important place in the upcoming missions such as the Euclid satellite \cite{Amendola:2016saw}, as it could reveal more on the nature of DE, viz., generically, modified gravity theories induce non-zero anisotropic stresses that lead to corresponding shear scalar evolutions, see, e.g., \cite{Pimentel89,Madsen88,Faraoni:2018qdr,Akarsu:2019pvi}. 

We introduce a generalization of the usual vacuum energy, called \textit{deformed vacuum energy}, which yields anisotropic pressure whilst preserving zero inertial mass density. It couples to the shear scalar in a unique way, such that they together emulate the canonical SFs with an arbitrary potential. This leads to the opportunity of reconsidering the cosmologies employing canonical SF, along with a bonus that the kinetic term of the SF is replaced by a new independent observable, the shear scalar.

\emph{Deformed vacuum energy} --
We begin with the locally rotationally symmetric (LRS) Bianchi I metric given by
\begin{equation}
\label{eq:metric}
 {\rm d}s^2 = -{\rm d} t^2+s^2\,\left[e^{\frac{4}{\sqrt{6}}\varphi}{\rm d}x^2+ e^{-\frac{2}{\sqrt{6}}\varphi}({\rm d}y^2+{\rm d}z^2)\right],
\end{equation}
which simply allows a different scale factor along one of the principal axes of the spatially flat RW metric, while preserving the isotropic spatial curvature \cite{Collins:1972tf,EllisRC,Ellis:1998ct}. $s\equiv v(t)^{1/3}$ is the mean scale factor with comoving volume scale factor $v(t)$, from which the average Hubble parameter is defined as $\mathcal{H}\equiv\frac{\dot{s}}{s}=\frac{1}{3}(H_x+2H_y)$, where $H_{i}$ $(i=x,y,z)$ are the directional Hubble parameters along the $x$-, $y$- and $z$-axes, and dot denotes the comoving time $t$ derivative. The term $\varphi$ is related to the shear scalar $\sigma^2=\frac{3}{2}(H_x-\mathcal{H})^2$, quantifying the anisotropic expansion, as $\sigma^2=\dot{\varphi}^2$. We use the geometrised units $c=1=8\pi G$.

In a generic inertial frame, the most general matter energy-momentum tensor accommodated by the metric \eqref{eq:metric} can be decomposed relative to a unique four-velocity $u^{\mu}$ ($u_{\mu}u^{\mu}=-1$ and $\nabla_{\nu}u^{\mu}u_{\mu}=0$) in the form
\begin{align}
T_{\mu\nu}=\rho u_{\mu}u_{\nu}+p_{\rm iso}\,h_{\mu\nu}+\pi_{\mu\nu}.
\label{eq:emt}
\end{align}
Here $\rho=\rho(u^{\mu})=T_{\mu\nu}u^{\mu}u^{\nu}$ is the  relativistic energy density relative to $u^{\mu}$, $p_{\rm iso}= \frac{1}{3}T_{\mu\nu}h^{\mu\nu}$ is the isotropic pressure and $\pi_{\mu\nu} = T_{\lambda\zeta}\,h^{\lambda}{}_{\langle
\mu}\,h^{\zeta}{}_{\nu \rangle}$ is the trace-free anisotropic pressure, where $h_{\mu\nu}=g_{\mu\nu}+u_{\mu} u_{\nu}$ ($g_{\mu\nu}$ being the metric tensor) is the projection tensor into the instantaneous rest frame of comoving observers. The set of equations arises from the twice-contracted Bianchi identities, which by means of Einstein field equations $G_{\mu\nu} =-T_{\mu\nu}$, implies the conservation equations. Projecting parallel and orthogonal to $u^{\mu}$, we obtain the energy and momentum conservation equations, correspondingly, 
\begin{eqnarray}
\dot\rho+\Theta(\rho+p_{\rm iso})+\sigma_{\mu\nu}\pi^{\mu\nu} =0, \label{eq:continuity}\\
\D^{\mu} p_{\rm iso}+(\rho+p_{\rm iso}+\pi^{\mu}_\mu)\dot{u}^{\mu}+\left(\div{\pi}\right)^{\mu}=0,
\label{eq:eulerani}
\end{eqnarray}
where $\Theta=\D^{\mu}u_{\mu}$ is the volume expansion rate, $\sigma_{\mu\nu}=\D_{\langle \mu}u_{\nu\rangle }$ is the shear tensor, and we used $\nabla_{\nu}u_{\mu}=\D_{\nu}u_{\mu}-\dot{u}_{\mu} u_{\nu}$ with $\D_{\nu} u_{\mu}= {\ts{1\over3}}\Theta h_{\mu\nu}+\sigma_{\mu\nu}$ \cite{EllisRC,Ellis:1998ct}. We note that $\dot{u}^{\mu}$ is the four acceleration, and thereby the multipliers $\rho+p_{\rm iso}+\pi^{\mu}_{\mu}$ for the spatial components in \eqref{eq:eulerani} define the inertial mass densities along the principal axes as $\rho_{{\rm inert},x}\equiv\rho+p_{\rm iso}+\pi^{1}_1$ and $\rho_{{\rm inert},y}=\rho_{{\rm inert},z}\equiv\rho+p_{\rm iso}+\pi^{2}_2$. Furthermore, we can  define an average inertial mass density as $\bar{\rho}_{\rm inert}\equiv\frac{1}{3}\left(\rho_{{\rm inert},x}+2\rho_{{\rm inert},y}\right)$ leading to
\begin{equation}
\begin{aligned}
\label{eq:inertnull}
\bar{\rho}_{\rm inert}=\rho+p_{\rm iso}+\frac{1}{3}\pi^{1}_1+\frac{2}{3}\pi^{2}_2.
  \end{aligned}
  \end{equation}
  As the pressure along $x$-axis is $p_{x}=p_{\rm iso}+\pi^{1}_1$ and the ones along $y$- and $z$-axes are $p_{y}=p_{z}=p_{\rm iso}+\pi^{2}_2$, we can write $p_{y}=p_{x}+\gamma\rho$ with $\gamma\rho\equiv\pi^{2}_2-\pi^{1}_1$ measuring the deviation of $p_{y}$ from $p_{x}$, so that \eqref{eq:inertnull} can be written as \begin{equation}
\label{eq:inertnull2}
\bar{\rho}_{\rm inert}=\rho+p_x+\frac{2}{3}\gamma\rho.
  \end{equation}
Staying loyal to the zero inertial mass density of the usual vacuum energy ($\rho_{\rm v}+p_{\rm v}=0$), we thus assume
\begin{equation}
\label{eq:inertnull3}
\bar{\rho}_{\rm inert}=0,
\end{equation}
leading to $p_{x}=-\rho-\frac{2}{3}\gamma\rho$, from which, we reach a particular kind of anisotropic stress;
\begin{equation}
\label{eq:emtfinal}
{T_{\mu}}^{\nu}={\textnormal{diag}}\left[-1,-1-\frac{2}{3}\gamma,-1+\frac{1}{3}\gamma,-1+\frac{1}{3}\gamma\right]\,\rho,
\end{equation}
which, henceforth, we call \textit{deformed vacuum energy} (dv).
Here, for convenience, we use the notation ${T_{\mu}}^{\nu}={\textnormal{diag}}\,[-1,w_x,w_y,w_z]\,\rho={\textnormal{diag}}\,[-1,w_x,w_x+\gamma,w_x+\gamma]\,\rho$ with $\gamma=w_y-w_x$ being the skewness parameter providing a measure for the anisotropy of the fluid.


This is a well behaved anisotropic generalization of the usual vacuum energy of quantum field theory, which is isotropic ($\gamma=0$). Such that, if we set a cosmic triad \cite{ArmendarizPicon:2004pm}, that is a set of three identical of them pointing mutually in orthogonal spatial directions, then these three resemble exactly the usual vacuum energy. Similarly, arbitrary number of them oriented in arbitrary directions would on average lead, stochastically, to the usual vacuum energy, cf. \cite{Golovnev:2008cf}. Besides, it does not represent any of the well known anisotropic sources such as vector fields, topological defects, etc. \cite{Barrow:1997sy}. For instance, the EoS of a vector field $A_{\mu}$ with a mass $m$, $ -w_x=w_y=w_z=\frac{\dot{A}^2-m^2A^2}{\dot{A}^2+m^2A^2}$ \cite{Koivisto:2008xf}, implies $w_x=-\gamma/2$ with $-2\leq\gamma\leq 2$ for $m^2\geq 0$, which do not satisfy \eqref{eq:emtfinal}. Similarly, the topological defects such as cosmic strings $\{w_x,\gamma\}=\{-1,1\}$, or domain walls $\{w_x,\gamma\}=\{0,-1\}$, do not satisfy \eqref{eq:emtfinal}.

\emph{Emulator for canonical scalar field} --
The Einstein field equations in the presence of the deformed vacuum energy \eqref{eq:emtfinal} for the simplest anisotropic background \eqref{eq:metric} can be given by the following set of equations:
\begin{align}
 3\mathcal{H}^2&=\frac{1}{2}\sigma^2+\rho_{\rm {dv}}\label{eq:rho2r}, \\
-2 \dot{\mathcal{H}}-3\mathcal{H}^2&=\frac{1}{2}\sigma^2-\rho_{\rm {dv}} \label{eq:pav}, \\
\dot{\sigma}+3\mathcal{H}\sigma&=-\sqrt{\frac{2}{3}}\gamma\rho_{\rm {dv}},\label{eq:shearprop}
\end{align}
which are the energy density \eqref{eq:rho2r}, average pressure \eqref{eq:pav} and shear propagation \eqref{eq:shearprop} equations, respectively. 
Comparing the density and average pressure equations
, we see that the shear scalar term $\sigma^2/2$ and the density of the deformed vacuum $\rho_{\rm dv}$ together resemble the canonical SF, namely, they appear as the kinetic term $\dot{\phi}^2/2$ and potential $V$ of a canonical SF, correspondingly. Further the shear propagation equation \eqref{eq:shearprop} resembles the scalar field (Klein-Gordon) equation. Thus, this system \eqref{eq:rho2r}-\eqref{eq:shearprop} has the same mathematical form of the standard isotropic Friedmann equations in the presence of a canonical SF
 \begin{align}
 3H^2&=\frac{1}{2}\dot{\phi}^2+V(\phi) \label{eq:phi1}, \\
-2 \dot H-3 H^2&=\frac{1}{2}\dot{\phi}^2-V(\phi) \label{eq:phi2}, \\
\ddot{\phi}+3 H\dot{\phi}&=-\frac{{\rm d} V}{{\rm d} \phi}
\label{eq:KG1},
\end{align}
under the following transformations:
\begin{equation}
\label{eq:transformations0}
\mathcal{H}\rightarrow H\,\, ,\,\,\sigma \rightarrow \dot{\phi}\,\, ,\,\,\rho_{\rm {dv}} \rightarrow V(\phi)\,\, ,\,\,\gamma\rightarrow\sqrt{\frac{3}{2}}\frac{1}{V} \frac{{\rm d} V}{{\rm d} \phi}.
\end{equation}
Accordingly, given that the energy density and pressure of a canonical SF are $\rho_{\phi}=\frac{1}{2}\dot{\phi}^2+V(\phi)$ and $p_{\phi}=\frac{1}{2}\dot{\phi}^2-V(\phi)$, if we define effective energy density and pressure as $\rho_{\rm eff}=\frac{1}{2}\sigma^2+\rho_{\rm dv}$ and $p_{\rm eff}=\frac{1}{2}\sigma^2-\rho_{\rm dv}$, correspondingly, we further have the transformations:
\begin{align}
\rho_{\rm eff}\rightarrow\rho_{\phi},\; p_{\rm eff}\rightarrow p_{\phi}\;\,{\rm and}\;\, w_{\rm eff}\equiv\frac{p_{\rm eff}}{\rho_{\rm eff}} \rightarrow
w_{\phi}\equiv\frac{p_{\phi}}{\rho_{\phi}}.
\label{eq:weff}
\end{align}
It is straightforward to see that as the Klein-Gordon equation \eqref{eq:KG1} leads to the continuity equation, $\dot{\rho}_{\phi}+3H(\rho_{\phi}+p_{\phi})=0$, for the SF, the shear propagation equation \eqref{eq:shearprop} leads to the continuity equation
\begin{equation}
\dot{\rho}_{\rm eff}+3\mathcal{H}\rho_{\rm eff}(1+w_{\rm eff})=0,
\label{eq:rhoeff}
\end{equation}
for the effective source defined from the cooperation of the deformed vacuum energy with the shear scalar---as long as the shear propagation equation is not altered by any other anisotropic contribution.

While the condition for a canonical SF to be able to drive accelerated expansion, $w_{\phi}<-1/3$, implies $\dot{\phi}^2<V$ (or $\dot{\phi}^2/2<\rho_{\phi}/3$), in our model, correspondingly, $w_{\rm eff}<-1/3$ implies $\sigma^2<\rho_{\rm dv}$ (or $\sigma^2/2<\rho_{\rm eff}/3$). On the other hand, to give rise to an accelerated expansion using SF, it is often required a flat potential satisfying $\dot{\phi}^2\ll V$, which leads to $w_{\phi}\simeq -1+\frac{2}{3}\epsilon$, where $\epsilon\ll 1$ is the so-called slow roll parameter defined as $\epsilon=\frac{1}{2}(\frac{1}{V}\frac{{\rm d} V}{{\rm d} \phi})^2$. Considering the relations given in \eqref{eq:KG1} and \eqref{eq:transformations0}, it turns out that the role of the slow-roll parameter is taken over by the skewness of the deformed vacuum energy as $\gamma^2/3\rightarrow \epsilon$ and hence one should require small anisotropy $\sigma^2\ll\rho_{\rm dv}$, which leads to $w_{\rm eff}\simeq -1+\frac{2}{9}\gamma^2$ with $|\gamma|\ll\sqrt{3}$. And, the role of the flatness of the potential (quantified by $\epsilon$) is taken over by the ratio-squared of the rate of change of the energy density of the deformed vacuum to the shear scalar, namely, $\epsilon\rightarrow \frac{\gamma^2}{3}=\frac{1}{2}\frac{\dot{\rho}_{\rm dv}^2}{\rho_{\rm dv}^2}\frac{1}{\sigma^2}$.

There is no-go theorem which forbids a single canonical SF (real SF $\dot{\phi}^2\geq0$ with a non-negative potential $V(\phi)\geq0$) to cross below the $w=-1$ boundary of the usual vacuum energy, viz., its EoS parameter is confined to the range $-1 \leq w_{\rm \phi}\leq 1$. In line with that, in our model, the non-negativity condition on the density of the deformed vacuum energy---as an actual physical source with negative density would be physically ill---($\rho_{\rm dv}\geq0$) along with that the shear scalar is non-negative definite by itself ($\sigma^2\geq0$) guarantee that $-1\leq w_{\rm eff}\leq 1$.

\emph{Cosmology with deformed vacuum energy} --
We proceed with an investigation of the cosmologies in the presence of the deformed vacuum. We consider the isotropic perfect fluids---representing usual cosmological fluids such as dust and radiation---described by $p_i/\rho_i=w_i=\rm const$ ($i$ stands for $i^{\rm th}$ fluid) for the other sources present along with the deformed vacuum energy in the Universe. As these are isotropic, these alter neither the form of the shear propagation equation \eqref{eq:shearprop} nor the features that arise from the deformed vacuum energy, see \eqref{eq:transformations0}-\eqref{eq:rhoeff}. 

Using ${\rm d}t=-\frac{{\rm d}z}{\mathcal{H}(1+z)}$, where  $z$ is the average redshift defined from the mean scale factor as $z=\frac{s(t=0)}{s(t)}-1$, we reach the following anisotropic Friedmann equation
\begin{align}
3\mathcal{H}^2=\sum_{i}\rho_{i0} (1+z)^{3(1+w_{ i})}+\rho_{\rm eff}\label{eq:q1},
\end{align}
where
\begin{align}
\rho_{\rm eff}=\rho_{\rm eff0}\,{\rm e}^{3\int (1+w_{\rm eff})\,{\rm d}\ln{(1+z)}}.
\label{eq:rhoeffz}
\end{align}
This is mathematically exactly the same with the usual Friedmann equation, but physically different. Note that here $\rho_{\rm eff}=\rho_{\sigma^2}+\rho_{\rm dv}$ consists of the energy density corresponding to the shear scalar, i.e., expansion anisotropy,
\begin{align}
\rho_{\sigma^2}\equiv\frac{\sigma^2}{2}=\frac{1+w_{\rm eff}}{2}\rho_{\rm eff}, \label{eq:sigmadef}
\end{align}
and the energy density of the deformed vacuum,
\begin{align}
\label{eq:rhodv}
\rho_{\rm dv}=\frac{1-w_{\rm eff}}{2}\rho_{\rm eff},
\end{align}
whose EoS parameter is skewed as
\begin{align}
\label{eq:sigmadef2}
\gamma=\frac{w_{\rm eff}'(1+z)-3(1-w_{\rm eff}^2)}{\sqrt{2+2w_{\rm eff}}(1-w_{\rm eff})}\sqrt{1+\frac{\sum_{i}\rho_{ i}}{\rho_{\rm eff}}},
\end{align}
where $^\prime$ denotes ${\rm d}/{\rm d}z$.

We see that the ratio of the energy density corresponding to the expansion anisotropy to that of the deformed vacuum energy is determined solely by $w_{\rm eff}$ as $\frac{\rho_{\sigma^2}}{\rho_{\rm dv}}= \frac{\Omega_{\sigma^2}}{\Omega_{\rm dv}}=\frac{1+w_{\rm eff}}{1-w_{\rm eff}}$. $\Omega=\rho/\rho_{\rm cr}$ (with $\rho_{\rm cr}=3\mathcal{H}^2$ being the critical energy density) is the density parameter of the component denoted by its subscript. Accordingly, if the energy density/EoS of a SF is given in terms of redshift, then one can straightforwardly study its correspondence via the transformations $\rho_{\phi}(z)\rightarrow\rho_{\rm eff}(z)$ or $w_{\phi}(z)\rightarrow w_{\rm eff}(z)$ [see \eqref{eq:weff}]---, upon first replacing the RW background by the Bianchi I (viz., $H\rightarrow \mathcal{H}$). If the potential, $V(\phi)$, described the SF is given, the same procedure can be utilized upon obtaining $\rho_{\phi}(z)$ from the exact solutions of the considered cosmological model. The emulator of the given SF will give exactly the same expansion history for the comoving volume element. Yet, each of the SF will correspond to a specific evolution of the shear scalar, which would allow, in principle, to observationally distinguish between these two corresponding models.

We continue the exploration of the model by focusing on DE. Dark energies described by a canonical SF with a non-trivial potential in which the field slowly rolls down is the epitome of quintessence models \cite{Tsujikawa:2010sc,Tsujikawa:2013fta}, and thereby the shear scalar-deformed vacuum energy alliance would also be. To begin with, in the case of emulating a SF with $w_{\phi}=\rm const$ leading to $\rho_{\phi}\propto (1+z)^{3(1+w_{\phi})}$ (the so called $w$CDM model but with a lower bound $w=-1$), the ratio $\frac{\Omega_{\sigma^2}}{\Omega_{\rm eff}}=\frac{1+w_{\rm eff}}{2}$ is a fixed value, so that, as $w_{\phi}\rightarrow w_{\rm eff}$, for the shear scalar we have $\sigma^2\propto (1+z)^{3(1+w_{\rm eff})}$ ---i.e., it tracks $\rho_{\rm eff}\propto (1+z)^{3(1+w_{\rm eff})}$---, which, in principle, is an observable and can be utilized to distinguish between our model and the usual $w$CDM based on RW background. As for the $w$CDM observations suggest $w\sim-1$, this implies $\sigma^2\sim \rm const$ along with $\Omega_{\sigma^2}$ increasing with the expansion of the Universe (anisotropization of the Universe as it expands) in contrast to the simple Bianchi I generalization of the standard $\Lambda$CDM (or of any isotropic DE) for which $\sigma^2\propto (1+z)^6$. The situation changes in the case of early DE which was suggested for addressing the so called Hubble tension as it increases the early expansion rate while leaving the later evolution of the Universe unaltered \cite{Poulin:2018cxd}. It behaves like a cosmological constant, before some critical redshift, $w\sim-1$ for $z<z_{\rm c}$, but its energy density then increases like that of radiation with the increasing redshift, $w\sim\frac{1}{3}$ for $z>z_{\rm c}$. Accordingly, if we emulate early DE, then $\sigma^2\sim\rm const$ and $\frac{\Omega_{\sigma^2}}{\Omega_{\rm eff}}\sim0$ (almost isotropic Universe) before some critical redshift, as $w_{\rm eff}\sim-1$ for $z<z_{\rm c}$, but afterwards shear scalar increases like the energy density of radiation and $\frac{\Omega_{\sigma^2}}{\Omega_{\rm eff}}\sim\frac{2}{3}$, as $w_{\rm eff}\sim\frac{1}{3}$ for $z>z_{\rm c}$. Thus, the emulator of a quintessence can be distinguished via the modified redshift dependence of the shear scalar.

A canonical SF with an exponential potential, $V(\phi)=V_0 e^{\lambda\phi}$ with $\lambda=\rm const$, corresponds to the case of the deformed vacuum energy with a constant skewness parameter, $\gamma=\rm const$. The corresponding transformation for which reads $\gamma\rightarrow\sqrt{\frac{3}{2}}\lambda$ from \eqref{eq:transformations0}. One of the two Swampland criteria on an effective field theory consistent with string theory is that given a point in field space, the derivative of the SF potential has to satisfy the lower bound $\frac{|{\rm d}V/{\rm d}\phi|}{V}>c\sim \mathcal{O}(1)$. For exponential potential of the form $V(\phi)=V_0 e^{\lambda\phi}$, $\lambda=c$. The dark energies that barely distinguished from the $\Lambda$ require $\lambda\leq0.1$ ($w_{\phi}\leq -0.998$) in great tension with the string Swampland criterion \cite{Heisenberg:2018yae}. On the other hand, because our model does not actually include SF, but emulates it, the swampland criteria do not apply anymore. The observational requirement $\lambda\leq0.1$ corresponds to $\gamma\leq0.12$, $w_{\rm eff}\leq-0.998$ and $\frac{\Omega_{\sigma^20}}{\Omega_{\rm dv0}}\leq0.001$ in our model, which, for $\Omega_{\rm eff0}\sim0.7$, leads to $\Omega_{\sigma^20}\lesssim7\times10^{-4}$ for the present-day density parameter corresponding to the expansion anisotropy, which, interestingly, matches the upper bounds from SN Ia observations \cite{Campanelli:2010zx,Wang:2017ezt}. Of course, whether the deformed vacuum energy could be obtained from a fundamental theory remains as an open question.

For a demonstration of how the model works and to give a guide to the values of its parameters, we continue with the emulator for one parameter extension of the $\Lambda$CDM---the so called $w$CDM model replacing the $\Lambda$ by a DE with $w=\rm const$---and compare with the simplest anisotropic extension of the $\Lambda$CDM (henceforth, $\Lambda$CDM$_{\sigma^2}$) \cite{Akarsu:2019pwn}. The $w$CDM for $-1\leq w \leq 1$ can be well described by a SF with a corresponding suitable potential \cite{Rubano:2001xi}. We present in Table \ref{tab:theory} the corresponding equations of the relevant parameters for this emulator (henceforth, dv$w$CDM$_{\sigma^2}$) obtained from \eqref{eq:q1}-\eqref{eq:sigmadef2} by following similar mathematical procedures used in \cite{Rubano:2001xi}, and, for a comparison, those of $\Lambda$CDM$_{\sigma^2}$. In the case of $\Lambda$CDM$_{\sigma^2}$, $w_{\rm eff}$ resembles a SF with constant potential, i.e., $\rho_{\rm dv}=\rm const$ and $\rho_{\sigma^2}\propto (1+z)^6$ (likewise the stiff fluid). In the case of  dv$w$CDM$_{\sigma^2}$, on the other hand, $w_{\rm eff}$ is a constant, corresponding to a SF of which the kinetic term and the potential have the same redshift dependence as $\rho_{\sigma^2}= \frac{1}{2}(1+w_{\rm eff}) \rho_{\rm eff}$ and $\rho_{\rm dv}= \frac{1}{2}(1-w_{\rm eff}) \rho_{\rm eff}$ with $\rho_{\rm eff}=\rho_{\rm eff0}(1+z)^{3(1+w_{\rm eff})}$.

\begin{table}[t!]\footnotesize
\caption{Equations for $\Lambda$CDM$_{\bf \sigma^2}$ and dv$w$CDM$_{\sigma^2}$ models.}
\label{tab:theory}
\scalebox{0.9}{
{\renewcommand{\arraystretch}{2}
\begin{tabular}{|c|c|c|}
\hline & $\Lambda$CDM$_{\sigma^2}$ & dv$w$CDM$_{\sigma^2}$  \\  \hline\hline
$\rho_{\rm eff}$ & $\rho_{\rm dv}+\rho_{\sigma^20}(1+z)^{6}$  & $\rho_{\rm eff0}(1+z)^{3(1+w_{\rm eff})}$   \\  \hline 
$w_{\rm eff}$ & $\frac{\rho_{\sigma^20}(1+z)^{6}-\rho_{\rm dv}}{\rho_{\sigma^20}(1+z)^{6}+\rho_{\rm dv}}$  & $\rm{const.}\geq-1$ \\  \hline \hline
 $\rho_{\sigma^2}$  & $\rho_{\sigma^20}(1+z)^{6}$ & $\frac{1}{2}(1+w_{\rm eff})\rho_{\rm eff0}(1+z)^{3(1+w_{\rm eff})}$ \\ \hline 
$\rho_{\rm dv}$  & $\rm const$ & $\frac{1}{2}(1-w_{\rm eff})\rho_{\rm eff0}(1+z)^{3(1+w_{\rm eff})}$ \\ \hline
$\gamma$ & $0$ & $-3\sqrt{\frac{1+w_{\rm eff}}{2}\left[1+\frac{\rho_{\rm m0}}{\rho_{\rm eff0}}(1+z)^{-3w_{\rm eff}}\right]}$ \\
\hline
\end{tabular}}}
\end{table}

$\Lambda$CDM$_{\sigma^2}$ model, which brings in the term $\rho_{\sigma^2}\propto (1+z)^{6}$ in the average expansion rate $\mathcal{H}$, is well constrained. Through this term, it is found in a recent study that $\Omega_{\sigma0}\lesssim10^{-3}$ from Hubble and Pantheon data, and, when the baryonic acoustic oscillations (BAO) and cosmic microwave background (CMB) data are included, $\Omega_{\sigma^20}\lesssim10^{-15}$, for which anisotropy becomes irrelevant to the matter-radiation equality redshift and the peak of the matter perturbations, but the CMB quadrupole temperature changes up to values beyond its actual value, viz., $\sim11\, \rm mK$. Besides, it was suggested that the anisotropy has no significant effect on the standard Big Bang Nucleosynthesis (BBN) provided that $\Omega_{\sigma^20}\lesssim10^{-23}$, for which anisotropy remains irrelevant to the CMB quadrupole temperature. On the other hand, in the dv$w$CDM$_{\sigma^2}$ model, the shear scalar tracks the deformed vacuum energy---both of which evolve as $(1+z)^{3(1+w_{\rm eff})}$---and hence, as like the DE in the usual $w$CDM, it would reach considerable values at late times only, and consequently the constraints on the anisotropy can be relaxed. Namely, in this case, the Universe anisotropizes as it expands, which implies that the expansion anisotropy would be irrelevant to the dynamics of the early Universe and the tight constraints on its present-day density parameter from its effect on the expansion rate on the comoving volume of the early Universe (e.g., from BBN) would be evaded. This relaxed amount of anisotropic expansion would allow us to manipulate the CMB quadrupole temperature on top of its statistical value. This is the observationally distinguishing feature of the dv$w$CDM$_{\sigma^2}$ model from the usual $w$CDM model.

\emph{Manipulating CMB quadrupole temperature} --
\label{sec:cmbquadrupole}
The observed quadrupole power spectrum of temperature fluctuations in the CMB (multipole $\ell=2$, the corresponding angular scale $\theta=\pi/2$) is $\Delta T_{\rm PLK} \approx 14 \,\mu K$ \cite{Ade:2013kta} lower than the $\Lambda$CDM predicted value, $\Delta T_{\rm st}\approx 34\, \mu K$ (or $\Delta T_{\rm st+variance} \approx 28 \,\mu K$  when the cosmic variance is included) \cite{Dodelson03,Campanelli:2006vb}. It is suggested in \cite{BeltranJimenez:2007rsj,Koivisto:2005mm,Campanelli:2006vb,Koivisto:2007bp,Rodrigues:2007ny,Koivisto:2008xf,Campanelli:2007qn,Campanelli2,Campanelli:2009tk,Cea:2019gnu} that this discrepancy can be addressed by an ellipsoidal expansion (within LRS Bianchi I spacetime) driven by an anisotropic DE. The evolution of the free streaming photon temperature in the $i^{\rm th}$ direction can be given as $T_{i}=T_0\frac{a_{{\rm i}0}}{a_{\rm i}}=T_0 e^{-\int H_i {\rm d} t}\simeq {T_0}-{T_0}\int H_i {\rm d} t$ ($i=x,y,z$) where $T_0 =2.7255 \pm 0.0006 \,{\rm K}$ \cite{fixsen09} is the present-day CMB monopole temperature \cite{Barrow:1997sy,Barrow:1985tda}. Thus, as $H_x=\mathcal{H}+\frac{2}{\sqrt{6}}\sigma$,  $H_y=\mathcal{H}-\frac{1}{\sqrt{6}}\sigma$, and $\Omega_{\sigma^2}=\frac{\sigma^2/2}{3\mathcal{H}^2}$, the difference between photon temperatures along the $x$- and $y$-axes since the recombination ($z_{\rm rec}=1100$) to the present time ($z=0$) due to the anisotropic expansion, $\Delta T_{\sigma^2}=T_x-T_y$, reads
\begin{equation}
\begin{aligned}
\Delta T_{\sigma^2}=T_0\int_{t_{\rm rec}}^{t_0}(H_x-H_y){\rm d}t
&=3T_0\int_0^{z_{\rm rec}}\sqrt{\Omega_{\sigma^2}}\,{\rm d}\ln (1+z).\nonumber
\label{deltaT}
\end{aligned}
\end{equation}
Accordingly, provided that the orientation of the expansion anisotropy is set properly, by means of $\Delta T_{\sigma^2}\approx 20 \,\mu K$ it is possible to reduce $\Delta T_{\rm st}\approx 34 \mu K$ in $\Lambda$CDM to the observed value $\Delta T_{\rm PLK} \approx 14 \mu K$ \cite{Ade:2013kta}. However, within $\Lambda$CDM$_{\sigma^2}$, it is not possible to have this reduction, since the upper limit $\Omega_{\sigma^20}\sim 10^{-23}$ from BBN allows only up to $\Delta T_{\sigma^2}\sim 1\,\mu K$ reduction \cite{Akarsu:2019pwn}. In our model (even in the simplest case, dv$w$CDM$_{\sigma^2}$), we are able to manipulate the evolution of $\Omega_{\sigma^2}$ so as to evade this limit on $\Omega_{\sigma^20}$ from BBN and manipulate $\Delta T_{\sigma^2}$ at the required amount. This can be done by demanding, for instance, from the simplest case dv$w$CDM$_{\sigma^2}$, to lead to $\Delta T_{\sigma^2}\sim 20 \mu K$ change on top of $\Delta T_{\rm st}\approx 34 \mu K$ via a suitably choosing, e.g., present-day value of the expansion anisotropy, or, in a robust way, e.g., by including $\Delta T_{\rm PLK}\approx 14 \mu K$ as a prior while modelling anisotropic distribution of the data in the sky.


\emph{Observational constraints} --  We perform a parameter estimation and provide observational constraints of the model-free parameters given in Table \ref{tab:theory}. In order to explore the parameter space, we make use of a modified version of a simple and fast Markov Chain Monte Carlo (MCMC) code, named SimpleMC \cite{Anze,Aubourg:2014yra}, that computes expansion rates and distances using the Friedmann equation. For the model dv$w$CDM$_{\sigma^2}$, the Friedmann equation \eqref{eq:q1} in the presence of radiation ($w_{\rm r}=\frac{1}{3}$) and dust (CDM+baryons) ($w_{\rm m}=0$) reads:
\begin{equation}
\begin{aligned}
\label{eq:fried}
\frac{\mathcal{H}^2}{\mathcal{H}_0^2}=\Omega_{\rm eff0} (1+z)^{3(1+w_{\rm eff})}+\Omega_{\rm m0}(1+z)^{3}+\Omega_{\rm r0}(1+z)^{4},\nonumber
\end{aligned}
\end{equation}
where $\Omega_{\rm eff0}=\Omega_{\sigma^20}+\Omega_{\rm {dv}0}$. The code uses a compressed version of the recent Planck data (PLK), a recent reanalysis of Type Ia supernova (SN) data, and high-precision Baryon Acoustic Oscillation measurements (BAO) at different redshifts up to $z=2.36$ \cite{Aubourg:2014yra}. For a detailed description about the data sets used see \cite{Aubourg:2014yra}. We also include a collection of currently available measurements on $H(z)$ from cosmic chronometers ($H$) (see \cite{Gomez-Valent:2018hwc} and refs. therein). See \cite{Padilla:2019mgi} for an extended review of cosmological parameter inference procedure. Throughout the analysis we assume flat priors over our sampling parameters: $\Omega_{{\rm m}0}=[0.05,0.5]$ for the dust density parameter today, $\Omega_{{\rm b}0} h_0^2=[0.02,0.025]$ for the physical baryon density parameter today and $h_0=[0.4,1.0]$ for the reduced Hubble constant, $h_0=H_0/100\, {\rm km\,s}^{-1}{\rm Mpc}^{-1}$. As our main purpose here is to demonstrate how dv$w$CDM$_{\sigma^2}$ works in comparison with the $\Lambda$CDM ($w_{\rm eff}=-1$) and $\Lambda$CDM$_{\sigma^2}$ models, rather than providing robust observational analyses, for the sake of obtaining tight constraints consistent with $\Delta T_ {\sigma^2}\sim20\mu K$, we take samples from the posterior distribution of the parameter-space by imposing the condition $\Omega_{\sigma^20}=4\times 10^{-21}$ for $\Lambda$CDM$_{\sigma^2}$ and $ 1+w_{\rm eff}=1.900\times 10^{-11}$ for dv$w$CDM$_{\sigma^2}$ both of them are identified by the angle brackets in Table \ref{tab:obs}. Table \ref{tab:obs} summarizes the observational constraints on the free parameters (as well as the derived parameters labelled by *) of these three models using the combined data sets PLK+BAO+SN+$H$.

\begin{table}[t!]\footnotesize
 \caption{Constraints ($68\%$ C.L.) on the parameters using the combined data sets PLK+BAO+SN+$H$. Along the analysis, free parameter $w_{\rm eff}=w_0$ is fixed to a certain value $\langle -1+1.900\times 10^{-11}\rangle$ to restrict our analysis to $\Delta T_{\sigma^2} \sim 20 \,\mu K$ region. Derived parameters are labeled with $^*$ and the chosen parameters are enclosed in angle brackets.}
 \label{tab:obs}
\scalebox{1}{
\begin{tabular}{lcccc}
\hline\hline
    & $\Lambda$CDM  & $\Lambda$CDM$_{\sigma^2}$ & dv$w$CDM$_{\sigma^2}$  \\ [2pt]
    \hline
$\mathcal{H}_0$ $[\rm km\,s^{-1}\,Mpc^{-1}]$                    & $68.20(49)$ &  $68.84(49)$ & $67.65(85)$ 
\\ [2pt]
$\Omega_{\rm m0}$      & $0.302(6)$ &  $0.298(6)$ & $0.307(8)$ 
\\ [2pt]
$\Omega_{\sigma^20}$ & 0 & $\langle4\times 10^{-21}\rangle$ & $6.58(8) ~[10^{-12}]$ 
\\ [2pt]
$\Omega_{{\rm eff0}}$  &$0.698(6)$ & $0.701(6)$ & $0.693(8)$ 
\\ [2pt]
$ 1+w_{\rm eff}$                & $\langle 0 \rangle $ & 0 & $\langle 1.900\times 10^{-11}\rangle$ 
\\ [2pt]
$\gamma_0^*$              & 0 & 0 & $-11.108(49) ~[10^{-6}]$ 
\\ [2pt]
$\Delta T_ {\sigma^2}^*$ $[\mu K]$ & 0 & $21.12(21)$ & $20.08(27)$ 
\\ [2pt]
$k_{\rm eq}^*$ $[\rm Mpc^{-1}]$            &$0.01024(7)$  & $0.01032(7)$  & $0.01026(8)$ 
\\ [2pt]
$\Omega_{\sigma^2}(z=z_{\rm BBN})^*$ & 0 &  $0.803(2)$ & $8.91(32)~[10^{-42}]$ 
\\ [2pt]
\hline\hline
\end{tabular}}
\end{table}

\begin{figure}[t!]\centering
\includegraphics[width=0.45\textwidth]{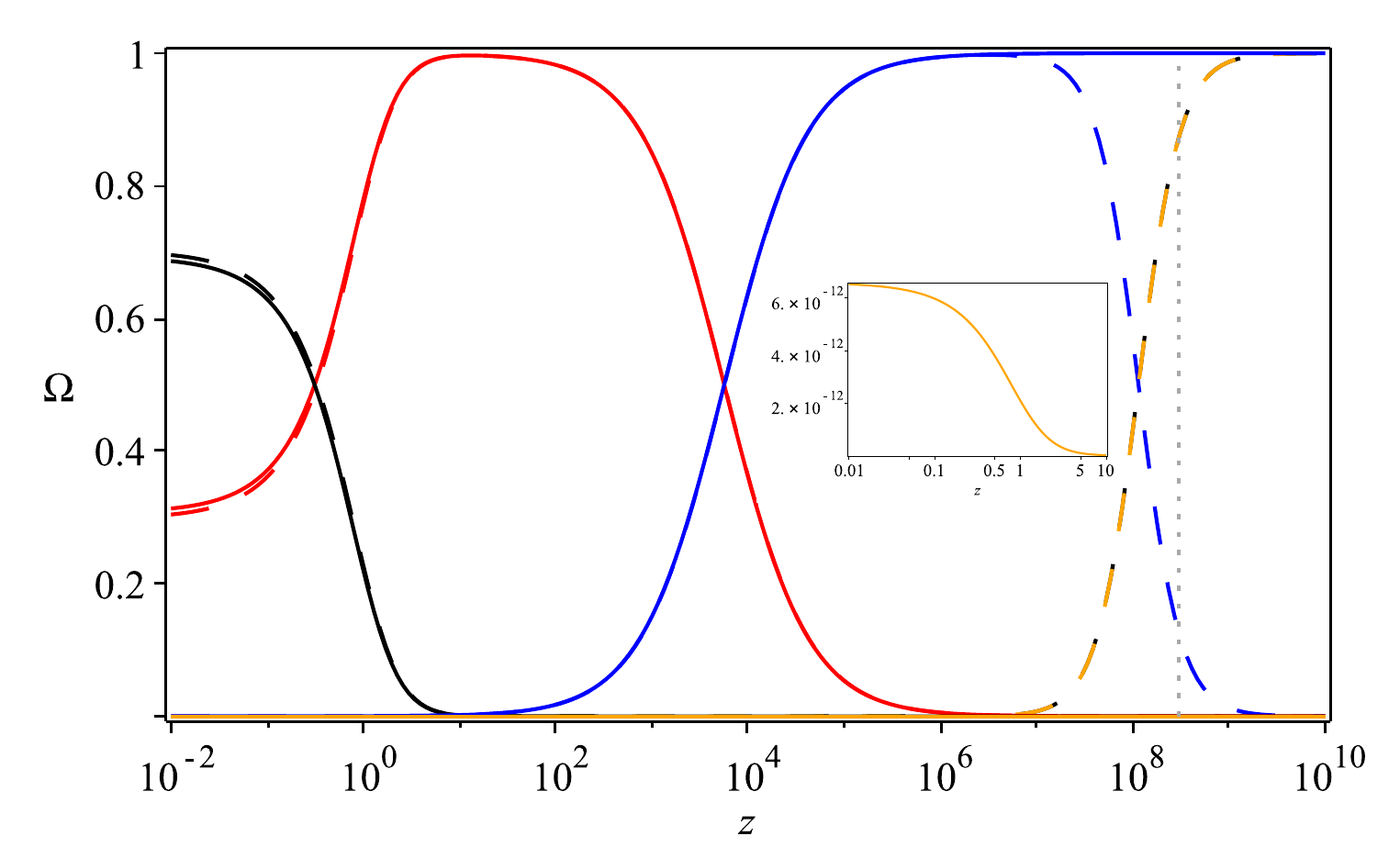}
\caption{$\Omega$ versus $z$ for $\Lambda$CDM$_{\bf \sigma^2}$ (dashed line) and dv$w$CDM$_{\sigma^2}$ (solid line) models using the mean values from Table \ref{tab:obs}. $\Omega_{\sigma^2}$, $\Omega_{\Lambda}$, $\Omega_{\rm m}$ and $\Omega_{\rm r}$ are colored by orange, black, red and blue, respectively. The vertical line represents the BBN epoch ($z_{\rm BBN}\sim3\times 10^8$).}
\label{fig:Omz}
\end{figure}

In Table \ref{tab:obs}, we observe that there exists no significant difference between the constraints on the parameters $\mathcal{H}_0$, $\Omega_{\rm m0}$ and $\Omega_{\rm eff0}$ of the models, and that the present-day density parameter corresponding to the anisotropic expansion, $\Omega_{\sigma^20}$, is of the order $\mathcal{O}(10^{-20})$ for $\Lambda$CDM$_{\sigma^2}$, and $\mathcal{O}(10^{-11})$ for dv$w$CDM$_{\sigma^2}$, which cannot be detected locally today---as they are much below the model independent upper bounds of order $\mathcal{O}(10^{-4})$. Further, we notice no significant difference between the constraints on $k_{\rm eq}=\frac{H_{\rm eq}}{1+z_{\rm eq}}$ (the wavenumber of a mode of density perturbations that enter the horizon at the radiation-matter transition, which is highly sensitive to the modifications to $\Lambda$CDM, and related to the dynamics of the Universe at matter-radiation equality redshift $z_{\rm eq}\sim3400$ larger than the recombination redshift $z_{\rm rec}\sim1100$ related to the CMB). All these imply that, when the evolution of the comoving volume element [viz., $\mathcal{H}(z)$] is considered, the $\Lambda$CDM$_{\sigma^2}$ and dv$w$CDM$_{\sigma^2}$ models are observationally indistinguishable from $\Lambda$CDM all the way to the matter-radiation transition epoch. Yet, both these can be distinguished from $\Lambda$CDM as they predict $\Delta T_ {\sigma^2}\sim20\mu K$, i.e., reduction of $\Delta T_{\rm st}\approx 34 \mu K$ in the $\Lambda$CDM to the observed value $\Delta T_{\rm PLK} \approx 14 \mu K$ \cite{Ade:2013kta}. However, the anisotropic expansion by this modification in the CMB quadrupole temperature does not spoil the successful description of the radiation dominated Universe (including standard BBN) only for the dv$w$CDM$_{\sigma^2}$. We see in Table \ref{tab:obs} and in Figure \ref{fig:Omz} that the expansion anisotropy dominates $80\%$ of the Universe at BBN epoch for $\Lambda$CDM$_{\sigma^2}$, while it is irrelevant to make any change on the standard BBN model for dv$w$CDM$_{\sigma^2}$. This implies in dv$w$CDM$_{\sigma^2}$ that it is not the BBN, but the quadrupole temperature putting the tightest constraints on the expansion anisotropy. Accordingly, while $\Lambda$CDM$_{\sigma^2}$ prohibits a significant modification in the CMB quadrupole temperature due to the tight BBN upper bound on the present-day expansion anisotropy, dv$w$CDM$_{\sigma^2}$ is able to manipulate it. Figure \ref{fig:Omz} is very demonstrative for the difference between these two anisotropic models. In $\Lambda$CDM$_{\sigma^2}$, as $\rho_{\sigma^2}\propto(1+z)^6$, the Universe isotropizes as it expands: The density parameter corresponding to expansion anisotropy $\Omega_{\sigma^2}$ rapidly increases---thereby the model deviates from $\Lambda$CDM---with increasing redshift, and eventually the expansion anisotropy dominates over the radiation and spoils the standard BBN (which must take place during radiation domination at $z\sim 3\times10^8$). In contrast, in dv$w$CDM$\sigma^2$, as $\rho_{\sigma^2}\sim \rm const$, the Universe anisotropizes as it expands: $\Omega_{\sigma^2}$ decreases---thereby the model approximates more and more to $\Lambda$CDM---with increasing redshift, vanishes almost completely before reaching to the redshift values relevant to the BBN processes---leaving standard BBN scenario unaltered---and further it completely vanishes in the beginning of the Universe.

\emph{Closing remarks} --  \label{sec:conc} We have introduced a generalization of the usual vacuum energy, called `deformed vacuum energy', which yields anisotropic pressure whilst preserving zero inertial mass density. It couples to the shear scalar in a unique way, such that they together emulate the canonical scalar field with an arbitrary potential. This leads to an interesting possibility of reconsidering the cosmologies employing a canonical SF. In this setup, the emulator of a given scalar field will give exactly the same expansion history for the comoving volume element, but, will distinguish (in principle, observationally as well) via the uniquely determined evolution of the expansion anisotropy depending on the potential of the considered scalar field. We further elaborate the aspects of replacing the quintessence---dark energy described by the canonical scalar field---by the deformed vacuum energy.

It would be interesting to extend our study to the inflationary cosmologies. It is straightforward to see that the Universe would be almost isotropic ($\Omega_{\sigma^2}\approx0$) during the quasi-de Sitter epoch (when $w_{\rm eff}\approx-1$) and then, while the Universe leaving this epoch, $w_{\rm eff}$ increases, so does $\Omega_{\sigma^2}$. This implies that emulators of the standard inflationary scenarios will generically predict an anisotropization process of the Universe by the end of inflation. This anisotropization (anisotropic hair) can occur in non-trivial ways, whence the reheating mechanisms and/or an actual scalar field is also included into the model.


Throughout the paper, we have considered the LRS Bianchi I spacetime (the simplest spatially flat anisotropic metric). Extending this work to Bianchi I or V (spatially open) spacetimes, in principle, would not change our results as these two are atypical in that they bring no restoring `force' term in the shear propagation equation \cite{Ellis:1998ct}, whereas one set of such terms come, in more complicated anisotropic spacetimes, 
anisotropic spatial curvature \cite{Barrow:1997sy}. For instance, the most general spatially flat (or open) anisotropic spacetimes, Bianchi VII$_0$ (or  VII$_h$), yield  anisotropic  spatial  curvature  that  mimics traceless  anisotropic  fluid. Thus, consideration of the deformed vacuum energy in more general anisotropic spacetimes would extend our approach presented here to a family of non-canonical scalar fields.

\emph{Acknowledgements} --  The authors thank to Shahin Sheikh-Jabbari for valuable discussions. \"{O}.A. acknowledges the support by the Turkish Academy of Sciences in scheme of the Outstanding Young Scientist Award  (T\"{U}BA-GEB\.{I}P). N.K. acknowledges the post-doctoral research support from the Istanbul Technical University (ITU). A.A.S. acknowledges funding from DST-SERB, Govt of India, under the project NO. MTR/20l9/000599. J.A.V. acknowledges the support provided by FOSEC SEP-CONACYT Investigaci\'on B\'asica  A1-S-21925, and UNAM-DGAPA-PAPIIT  IA102219.

\end{document}